# Computing All-Pairs Shortest Paths
# by Leveraging Low Treewidth

**Léon Planken**                                          L.R.PLANKEN@TUDELFT.NL
**Mathijs de Weerdt**                                      M.M.DEWEERDT@TUDELFT.NL
*Faculty of EEMCS, Delft University of Technology,*
*Delft, The Netherlands*

**Roman van der Krogt**                                    ROMAN@4C.UCC.IE
*Cork Constraint Computation Centre,*
*University College Cork, Cork, Ireland*

## Abstract

We present two new and efficient algorithms for computing all-pairs shortest paths. The algorithms operate on directed graphs with real (possibly negative) weights. They make use of directed path consistency along a vertex ordering $d$. Both algorithms run in $\mathcal{O}\left(n^2 w_d\right)$ time, where $w_d$ is the graph width induced by this vertex ordering. For graphs of constant treewidth, this yields $\mathcal{O}\left(n^2\right)$ time, which is optimal. On chordal graphs, the algorithms run in $\mathcal{O}\left(nm\right)$ time. In addition, we present a variant that exploits graph separators to arrive at a run time of $\mathcal{O}\left(nw_d^2 + n^2 s_d\right)$ on general graphs, where $s_d \leq w_d$ is the size of the largest minimal separator induced by the vertex ordering $d$. We show empirically that on both constructed and realistic benchmarks, in many cases the algorithms outperform Floyd–Warshall's as well as Johnson's algorithm, which represent the current state of the art with a run time of $\mathcal{O}\left(n^3\right)$ and $\mathcal{O}\left(nm + n^2 \log n\right)$, respectively. Our algorithms can be used for spatial and temporal reasoning, such as for the Simple Temporal Problem, which underlines their relevance to the planning and scheduling community.

## 1. Introduction

Finding shortest paths is an important and fundamental problem in communication and transportation networks, circuit design, bioinformatics, Internet node traffic, social networking, and graph analysis in general—e.g. for computing betweenness (Girvan & Newman, 2002)—and is a sub-problem of many combinatorial problems, such as those that can be represented as a network flow problem. In particular, in the context of planning and scheduling, finding shortest paths is important to solve a set of binary linear constraints on events, i.e. the Simple Temporal Problem (STP; Dechter, Meiri, & Pearl, 1991). The STP in turn appears as a sub-problem to the NP-hard Temporal Constraint Satisfaction Problem (TCSP; Dechter et al., 1991) and Disjunctive Temporal Problem (DTP; Stergiou & Koubarakis, 2000), which are powerful enough to model e.g. job-shop scheduling problems. The shortest path computations in these applications can account for a significant part of the total run time of a solver. Thus, it is hardly surprising that these topics have received substantial interest in the planning and scheduling community (Satish Kumar, 2005; Bresina, Jónsson, Morris, & Rajan, 2005; Rossi, Venable, & Yorke-Smith, 2006; Shah & Williams, 2008; Conrad, Shah, & Williams, 2009).

Instances of the STP, called Simple Temporal Networks (STNs), have a natural representation as directed graphs with real edge weights. Recently, there has been specific interest in STNs stemming from hierarchical task networks (HTNs; Castillo, Fernández-Olivares, & González, 2006; Bui & Yorke-Smith, 2010). These graphs have the "sibling-restricted" property: each task, represented by a pair of vertices, is connected only to its sibling tasks, its parent or its children. In these graphs the number of children of a task is restricted by a constant *branching factor*, and therefore the resulting STNs also have a tree-like structure.

The canonical way of solving an STP instance (Dechter et al., 1991) is by computing all-pairs shortest paths (APSP) on its STN, thus achieving full path consistency. For graphs with $n$ vertices and $m$ edges, this can be done in $\mathcal{O}(n^3)$ time with the Floyd–Warshall algorithm (Floyd, 1962), based on Warshall's (1962) formulation of efficiently computing the transitive closure of Boolean matrices. However, the state of the art for computing APSP on sparse graphs is an algorithm based on the technique originally proposed by Johnson (1977), which does some preprocessing to allow $n$ runs of Dijkstra's (1959) algorithm. Using a Fibonacci heap (Fredman & Tarjan, 1987), the algorithm runs in $\mathcal{O}(n^2 \log n + nm)$ time. In the remainder of this paper, we refer to this algorithm as Johnson.

In this paper we present two new algorithms for APSP with real edge weights (in Section 3). One algorithm, dubbed Chleq–APSP, is based on a point-to-point shortest path algorithm by Chleq (1995); the other, named Snowball, is similar to Planken, de Weerdt, and van der Krogt's (2008) algorithm for enforcing *partial* (instead of full) path consistency (P³C). These new algorithms advance the state of the art in computing APSP. In graphs of constant treewidth, such as sibling-restricted STNs based on HTNs with a constant branching factor, the run time of both algorithms is bounded by $\mathcal{O}(n^2)$, which is optimal since the output is $\Theta(n^2)$. In addition to these STNs, examples of such graphs of constant treewidth are outerplanar graphs, graphs of bounded bandwidth, graphs of bounded cutwidth, and series-parallel graphs (Bodlaender, 1986).

When Chleq–APSP and Snowball are applied to chordal graphs, they have a run time of $\mathcal{O}(nm)$, which is a strict improvement over the state of the art (Chaudhuri & Zaroliagis, 2000, with a run time of $\mathcal{O}(nmw_d^2)$; $w_d$ is defined below). Chordal graphs are an important subset of general sparse graphs: interval graphs, trees, $k$-trees and split graphs are all special cases of chordal graphs (Golumbic, 2004). Moreover, any graph can be made chordal using a so-called triangulation algorithm. Such an algorithm operates by eliminating vertices one by one, connecting the neighbours of each eliminated vertex and thereby inducing cliques in the graph.

The *induced width* $w_d$ of the vertex ordering $d$ is defined to be equal to the cardinality of the largest such set of neighbours encountered. The upper bound of the run time of both proposed algorithms on these general graphs, $\mathcal{O}(n^2 w_d)$, depends on this induced width. Finding a vertex ordering of minimum induced width, however, is an NP-hard problem (Arnborg, Corneil, & Proskurowski, 1987). This minimum induced width is the tree-likeness property of the graph mentioned above, i.e. the *treewidth*, denoted $w^*$. In contrast, the induced width is not a direct measure of the input (graph), so the bound of $\mathcal{O}(n^2 w_d)$ is not quite proper. Still, it is better than the bound on Johnson if $w_d \in o(\log n)$.[1]

---

1. We prefer to write $x \in o(f(n))$ instead of the more common $x = o(f(n))$. Formally, the right-hand side represents the set of all functions that grow strictly slower than $f(n)$, and the traditional equality in fact only works in one direction (see also Graham, Knuth, & Patashnik, 1989, Section 9.2).

To see this, note that the bound on `Johnson` is never better than $\mathcal{O}\left(n^2 \log n\right)$, regardless of the value of $m$.

In this paper, we also present a variant of `Snowball` that exploits graph separators and attains an upper bound on the run time of $\mathcal{O}\left(nw_d^2 + n^2 s_d\right)$. This upper bound is even better than the one for the two other new algorithms, since $s_d \leq w_d$ is the size of the largest minimal separator induced by the vertex ordering $d$. While theoretical bounds on the run time usually give a good indication of the performance of algorithms, we see especially for this last variant that they do not always predict which algorithm is best in which settings. In Section 4, therefore, we experimentally establish the computational efficiency of the proposed algorithms on a wide range of graphs, varying from random scale-free networks and parts of the road network of New York City, to STNs generated from HTNs and job-shop scheduling problems.

Below, we first give a more detailed introduction of the required concepts, such as induced width, chordal graphs and triangulation, after which we present the new algorithms and their analysis.

## 2. Preliminaries

In this section, we briefly introduce the algorithm that enforces directed path consistency (DPC) and how to find a vertex ordering required for this algorithm. We then present our algorithms for all-pairs shortest paths, all of which require enforcing DPC (or a stronger property) as a first step. In our treatment, we assume the weights on the edges in the graph are real and possibly negative.

### 2.1 Directed Path Consistency

Dechter et al. (1991) presented `DPC`, included here as Algorithm 1, as a way check whether an STP instance is consistent.[2] This is equivalent to checking that the graph does not contain a negative cycle (a closed path with negative total weight). The algorithm takes as input a weighted directed graph $G = \langle V, E \rangle$ and a vertex ordering $d$, which is a bijection between $V$ and the natural numbers $\{1, \ldots, n\}$. In this paper, we simply represent the $i$th vertex in such an ordering as the natural number $i$. The (possibly negative) weight on the arc from $i$ to $j$ is represented as $w_{i \to j} \in \mathbb{R}$. Our shorthand for the existence of an arc between these vertices, in either direction, is $\{i, j\} \in E$. Finally, we denote by $G_k$ the graph induced on vertices $\{1, \ldots, k\}$; likewise, for a set of vertices $V' \subseteq V$, $G_{V'}$ denotes the graph induced on $V'$. So, in particular, $G_V = G_n = G$.

In iteration $k$, the algorithm adds edges (in line 5) between all pairs of lower-numbered neighbours $i, j$ of $k$, thus triangulating the graph. Moreover, in lines 3 and 4, it updates the edge between $i$ and $j$ with the weight of the paths $i \to k \to j$ and $j \to k \to i$, if shorter. Consequently, for $i < j$, a defining property of `DPC` is that it ensures that $w_{i \to j}$ is no higher than the total weight of any path from $i$ to $j$ that consists only of vertices outside $G_j$ (except for $i$ and $j$ themselves). This implies in particular that after running `DPC`, $w_{1 \to 2}$ and $w_{2 \to 1}$ are labelled by the shortest paths between vertices 1 and 2.

---

2. Note that other algorithms—such as `Bellman-Ford`—can be used for this purpose as well, and usually perform better in practice.

---

**Algorithm 1:** DPC (Dechter et al., 1991)

**Input**: Weighted directed graph $G = \langle V, E \rangle$; vertex ordering $d : V \to \{1, \ldots, n\}$
**Output**: DPC version of $G$, or INCONSISTENT if $G$ contains a negative cycle

**1 for** $k \leftarrow n$ **to** 1 **do**
**2**     **forall** $i < j < k$ **such that** $\{i, k\}, \{j, k\} \in E$ **do**
**3**        $w_{i \to j} \leftarrow \min \{w_{i \to j}, w_{i \to k} + w_{k \to j}\}$
**4**        $w_{j \to i} \leftarrow \min \{w_{j \to i}, w_{j \to k} + w_{k \to i}\}$
**5**        $E \leftarrow E \cup \{\{i, j\}\}$
**6**        **if** $w_{i \to j} + w_{j \to i} < 0$ **then**
**7**           **return** INCONSISTENT
**8**        **end**
**9**     **end**
**10 end**
**11 return** $G = \langle V, E \rangle$

---

The run time of DPC depends on a measure $w_d$ called the *induced width* relative to the ordering $d$ of the vertices. Dechter et al. (1991) define this induced width of a vertex ordering $d$ procedurally to be exactly the highest number of neighbours $j$ of $k$ with $j < k$ encountered during the DPC algorithm. This includes neighbours in the original graph (i.e. $\{j, k\} \in E$) as well as vertices that became neighbours through edges added during an earlier iteration of the algorithm. However, the definition can be based on just the original graph and the vertex ordering, by making use of the following result.

**Proposition 1.** *Suppose that $G = \langle V, E \rangle$ is an undirected graph and $d : V \to \{1, \ldots, n\}$ (where $d$ is a bijection) is a vertex ordering. Suppose further that we are given $n$ sets of edges $E'_k$ for $1 \le k \le n$, defined as follows:*

$$E'_k = \left\{ \{j, k\} \subseteq V \mid j < k \land \exists path\ from\ k\ to\ j\ in\ G_{\{j\} \cup \{k, k+1, \ldots, n\}} \right\}$$

*Then, $E'_k$ is exactly the set of edges visited during iteration $k$ of DPC.*

*Proof.* Note that by definition, each set $E'_k$ is a superset of the original edges between vertex $k$ and its lower-numbered neighbours. We use this fact to prove the equivalence by induction.

The equivalence holds for the first iteration $k = n$, because $E'_n$ is exactly the set of original edges between vertex $n$ and its lower-numbered neighbours, and there are no earlier iterations during which DPC may have added other edges $\{j, k\}$ with $j < k$. Now, assuming that the equivalence holds for all sets $E'_\ell$ with $\ell > k$, we show that it also holds for $E'_k$. For this inductive case, we prove both inclusion relations separately.

($\supseteq$) To reach a contradiction, assume that there exists some edge $\{j, k\} \notin E'_k$, with $j < k$, which is visited by DPC during iteration $k$. Because $E'_k$ includes the original edges between $k$ and lower-numbered neighbours, this must be a new edge added during some earlier iteration $\ell > k$, so there must exist edges $\{j, \ell\}, \{k, \ell\} \in E'_\ell$. By the induction hypothesis, $j$ and $k$ are therefore connected in the induced subgraph $G_{\{j, k\} \cup \{\ell, \ell+1, \ldots, n\}}$. But then they

356

must also be connected in the larger subgraph $G_{\{j\}\cup\{k,k+1,...n\}}$ and thus by definition be included in $E'_k$: a contradiction.

($\subseteq$) Assume, again for reaching a contradiction, that there exists some edge $\{j,k\} \in E'_k$ not part of $E$ during iteration $k$ of DPC and therefore not visited by the algorithm. Clearly, $\{j,k\}$ cannot have been one of the original edges. By definition of $E'_k$ there must therefore exist a path with at least one intermediate vertex from $j$ to $k$ in the induced subgraph $G_{\{j\}\cup\{k,k+1,...n\}}$. Let $\ell$ be the lowest-numbered vertex other than $j$ and $k$ on this path; we have that $\ell > k > j$. Then, by the induction hypothesis, there must exist edges $\{j,\ell\}, \{k,\ell\} \in E'_\ell$, both of which were visited by DPC during iteration $\ell$. Once more, we reach a contradiction, since DPC must have added $\{j,k\}$ to $E$ during iteration $\ell > k$. $\qquad\square$

We now formally define the induced width as follows, and conclude with Proposition 1 that this is equivalent to the original procedural definition.

**Definition 1.** *Given an undirected graph $G = \langle V, E \rangle$, a vertex ordering $d$, and $n$ sets of edges $E'_k$ as in Proposition 1, the* induced width $w_d$ *of $G$ (relative to $d$) is the following measure:*

$$w_d = \max_{k \in V} \left| E'_k \right|$$

It follows that the run time of DPC is not a property of the graph per se; rather, it is dependent on both the graph and the vertex ordering used. With a careful implementation, DPC's time bound is $\mathcal{O}\left(nw_d^2\right)$ if this ordering is known beforehand.

The edges added by DPC are called *fill edges* and make the graph *chordal* (sometimes also called triangulated). Indeed, DPC differs from a triangulation procedure only by its manipulation of the arc weights. In a chordal graph, every cycle of length four or more has an edge joining two vertices that are not adjacent in the cycle. By Definition 1, the number of edges in such a chordal graph, denoted by $m_c \geq m$, is $\mathcal{O}\left(nw_d\right)$. We now give the formal definitions of these concepts.

**Definition 2.** *Given a graph $G = \langle V, E \rangle$ and a set $\left\{v^1, v^2, \ldots, v^k\right\} \subseteq V$ of vertices that form a cycle in $G$, a* chord *of this cycle is an edge between non-adjacent vertices in this cycle, i.e. an edge $\left\{v^i, v^j\right\} \in E$ with $1 < j - i < k - 1$. A graph $G = \langle V, E \rangle$ is called* chordal *if all cycles of size larger than 3 have a chord.*

**Definition 3.** *Given a graph $G = \langle V, E \rangle$, a* triangulation $T$ *of $G$, with $T \cap E = \varnothing$, is a set of edges such that $G' = \langle V, E \cup T \rangle$ is chordal. These edges are called* fill edges. *$T$ is a* minimal triangulation *of $G$ if there exists no proper subset $T' \subset T$ such that $T'$ is a triangulation of $G$.*

## 2.2 Finding a Vertex Ordering

In principle, DPC can use any vertex ordering to make the graph both chordal and directionally path-consistent. However, since the vertex ordering defines the induced width, it directly influences the run time and the number of edges $m_c$ in the resulting graph. As mentioned in the introduction, finding an ordering $d$ with minimum induced width $w_d = w^*$, and even just determining the treewidth $w^*$, is an NP-hard problem in general. Still, the class of constant-treewidth graphs can be recognised, and optimally triangulated, in $\mathcal{O}\left(n\right)$ time (Bodlaender, 1996). If $G$ is already chordal, we can find a *perfect ordering* (resulting in no fill edges) in $\mathcal{O}(m)$ time, using e.g. *maximal cardinality search* (MCS; Tarjan & Yannakakis, 1984). This perfect ordering is also called a simplicial ordering, because every vertex $k$ together with its lower-numbered neighbours in the ordering induces a clique (simplex) in the subgraph $G_k$. This implies the following (known) result, relating induced width and treewidth to the size of the largest clique in $G$.

**Proposition 2.** *If a graph $G$ is chordal, the size of its largest clique is exactly $w^* + 1$. If a non-chordal graph $G$ is triangulated along a vertex ordering $d$, yielding a chordal graph $G'$, the size of the largest clique in $G'$ is exactly $w_d + 1$. The treewidth of $G'$ equals $w_d$ and is an upper bound for the treewidth of the original graph $G$: $w^* \leq w_d$.*

For general graphs, various heuristics exist that often produce good results. We mention here the minimum degree heuristic (Rose, 1972), which in each iteration chooses a vertex of lowest degree. Since the ordering produced by this heuristic is not fully known before DPC starts but depends on the fill edges added, an adjacency-list-based implementation will require another $\mathcal{O}(\log n)$ factor in DPC's time bound. However, for our purposes in this article, we can afford the comfort of maintaining an adjacency matrix, which yields bounds of $\mathcal{O}(n^2 + nw_d^2)$ time and $\mathcal{O}(n^2)$ space.

## 3. All-Pairs Shortest Paths

Even though, to the best of our knowledge, a DPC-based APSP algorithm has not yet been proposed, algorithms for computing single-source shortest paths (SSSP) based on DPC can be obtained from known results in a relatively straightforward manner. Chleq (1995) proposed a point-to-point shortest path algorithm that with a trivial adaptation computes SSSP; Planken, de Weerdt, and Yorke-Smith (2010) implicitly also compute SSSP as part of their IPPC algorithm. These algorithms run in $\mathcal{O}(m_c)$ time and thus can simply be run once for each vertex to yield an APSP algorithm with $\mathcal{O}(nm_c) \subseteq \mathcal{O}(n^2 w_d)$ time complexity. Below, we first show how to adapt Chleq's algorithm to compute APSP; then, we present a new, efficient algorithm named Snowball that relates to Planken et al.'s (2008) P³C.

### 3.1 Chleq's Approach

Chleq's (1995) point-to-point shortest path algorithm was simply called Min−path and computes the shortest path between two arbitrary vertices $s, t \in V$ in a directionally path-consistent graph $G$. It is reproduced here as Algorithm 2 and can be seen to run in $\mathcal{O}(m_c)$ time because each edge is considered at most twice. The shortest distance from the source vertex $s$ is maintained in an array $D$; the algorithm iterates downward from $s$ to 1 and then upward from 1 to $t$, updating the distance array when a shorter path is found.

Since the sink vertex $t$ is only used as a bound for the second loop, it is clear that $D$ actually contains shortest distances between all pairs $(s, t')$ with $t' \leq t$. Therefore, we can easily adapt this algorithm to compute SSSP within the same $\mathcal{O}(m_c)$ time bound by setting $t = n$ and returning the entire array $D$ instead of just $D[t]$. We call the result Chleq−APSP, included as Algorithm 3, which calls this SSSP algorithm (referred to as Min−paths) $n$ times to compute all-pairs shortest paths in $\mathcal{O}(nm_c) \subseteq \mathcal{O}(nw_d^2)$ time.

---

**Algorithm 2:** Min–path (Chleq, 1995)

---

**Input**: Weighted directed DPC graph $G = \langle V, E \rangle$;
       (arbitrary) source vertex $s$ and destination vertex $t$

**Output**: Distance from $s$ to $t$, or INCONSISTENT if $G$ contains a negative cycle

---

**1** $\forall i \in V : D[i] \leftarrow \infty$

**2** $D[s] \leftarrow 0$

**3** **for** $k \leftarrow s$ **to** 1 **do**

**4**    **forall** $j < k$ **such that** $\{j, k\} \in E$ **do**

**5**    |   $D[j] \leftarrow \min\{D[j], D[k] + w_{k \to j}\}$

**6**    **end**

**7** **end**

**8** **for** $k \leftarrow 1$ **to** $t$ **do**

**9**    **forall** $j > k$ **such that** $\{j, k\} \in E$ **do**

**10**    |   $D[j] \leftarrow \min\{D[j], D[k] + w_{k \to j}\}$

**11**    **end**

**12** **end**

**13** **return** $D[t]$

---

**Algorithm 3:** Chleq–APSP

---

**Input**: Weighted directed graph $G = \langle V, E \rangle$; vertex ordering $d : V \to \{1, \ldots, n\}$

**Output**: Distance matrix $D$, or INCONSISTENT if $G$ contains a negative cycle

---

**1** $G \leftarrow \mathsf{DPC}(G, d)$

**2** **return** INCONSISTENT **if** DPC did

**3** **for** $i \leftarrow 1$ **to** $n$ **do**

**4**    |   $D[i][*] \leftarrow \mathsf{Min\text{-}paths}(G, i)$

**5** **end**

**6** **return** $D$

---

---

**Algorithm 4:** Snowball

    **Input**: Weighted directed graph $G = \langle V, E \rangle$; vertex ordering $d : V \to \{1, \ldots, n\}$
    **Output**: Distance matrix $D$, or INCONSISTENT if $G$ contains a negative cycle

**1**   $G \leftarrow \mathsf{DPC}(G, d)$
**2**   **return** INCONSISTENT **if** DPC did

**3**   $\forall i, j \in V : D[i][j] \leftarrow \infty$
**4**   $\forall i \in V : D[i][i] \leftarrow 0$
**5**   **for** $k \leftarrow 1$ **to** $n$ **do**
**6**      **forall** $j < k$ such that $\{j, k\} \in E$ **do**
**7**          **forall** $i \in \{1, \ldots, k-1\}$ **do**
**8**              $D[i][k] \leftarrow \min \{D[i][k], D[i][j] + w_{j \to k}\}$
**9**              $D[k][i] \leftarrow \min \{D[k][i], w_{k \to j} + D[j][i]\}$
**10**          **end**
**11**      **end**
**12**   **end**

**13**   **return** $D$

---

## 3.2 The Snowball Algorithm

In this section, we present an algorithm that computes APSP (or full path-consistency), dubbed Snowball and included as Algorithm 4, that has the same asymptotic worst-case time bounds as Chleq–APSP but requires strictly less computational work.

Like Chleq–APSP, this algorithm first ensures that the input graph is directionally path-consistent. The idea behind the algorithm is then that we grow, during the execution of the outermost loop, a clique $\{1, \ldots, k\}$ of computed (shortest) distances, one vertex at a time, starting with the trivial clique consisting of just vertex 1; while DPC performed a backward sweep along $d$, Snowball iterates in the other direction. When adding vertex $k$ to the clique, the two inner loops ensure that we compute the distances between $k$ and all vertices $i < k$. This works because we know by DPC that for any such pair $(i, k)$, there must exist a shortest path from $i$ to $k$ of the form $i \to \cdots \to j \to k$ (and vice versa), such that $\{j, k\} \in E$ with $j < k$ is an edge of the chordal graph. This means that the algorithm only needs to "look down" at vertices $i, j < k$, and it follows inductively that $D[i][j]$ and $D[j][i]$ are guaranteed to be correct from an earlier iteration.

The name of our algorithm derives from its "snowball effect": the clique of computed distances grows quadratically during the course of its operation. A small example of the operation of Snowball is given in Figure 1. Originally, the graph contained a shortest path 4–7–6–2–5–1–3. Dashed edges have been added by DPC, and the path 4–2–1–3 is now also a shortest path; in particular, $w_{4 \to 2}$ holds the correct value. This snapshot is taken for $k = 4$; the shaded vertices 1–3 have already been visited and shortest distances $D[i][j]$ have been computed for all $i, j \leq 3$. Then, during the iteration $k = 4$, for $j = 2$ and $i = 3$, the algorithm sets the correct weight of $D[4][3]$ by taking the sum $w_{4 \to 2} + D[2][3]$.

**Theorem 3.** *Algorithm 4 (*Snowball*) correctly computes all-pairs shortest paths in $\mathcal{O}\left(nm_c\right) \subseteq \mathcal{O}\left(n^2 w_d\right)$ time.*

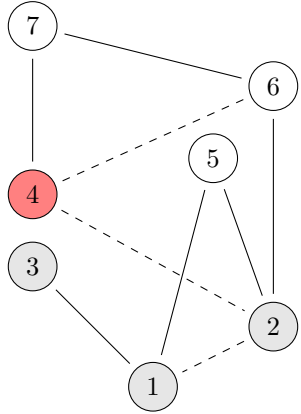

Figure 1: Snapshot ($k = 4$) of a graph during the operation of Snowball.

*Proof.* The proof is by induction. After enforcing DPC, $w_{1\to2}$ and $w_{2\to1}$ are labelled by the shortest distances between vertices 1 and 2. For $k = 2$ and $i = j = 1$, the algorithm then sets $D[1][2]$ and $D[2][1]$ to the correct values.

Now, assume that $D[i][j]$ is set correctly for all vertices $i, j < k$. Let $\pi : i = v_0 \to v_1 \to \cdots \to v_{\ell-1} \to v_\ell = k$ be a shortest path from $i$ to $k$, and further let $h_{\max} = \arg\max_{h \in \{0,1,\dots,\ell\}} \{v_h \in \pi\}$. By DPC, if $0 < h_{\max} < \ell$, there exists a path of the same weight where a shortcut $v_{h_{\max}-1} \to v_{h_{\max}+1}$ is taken. This argument can be repeated to conclude that there must exist a shortest path $\pi'$ from $i$ to $k$ that lies completely in $G_k$ and, except for the last arc, in $G_{k-1}$. Thus, by the induction hypothesis and the observation that the algorithm considers all arcs from the subgraph $G_{k-1}$ to $k$, $D[i][k]$ is set to the correct value. An analogous argument holds for $D[k][i]$.

With regard to the algorithm's time complexity, note that the two outermost loops together result in each of the $m_c$ edges in the chordal graph being visited exactly once. The inner loop always has fewer than $n$ iterations, yielding a run time of $\mathcal{O}(nm_c)$ time. From the observation above that $m_c \leq nw_d$, we can also state a looser time bound of $\mathcal{O}(n^2 w_d)$. $\square$

We now briefly discuss the consequences for two special cases: graphs of constant treewidth and chordal graphs. For chordal graphs, which can be recognised in $\mathcal{O}(m)$ time, we can just substitute $m$ for $m_c$ in the run-time complexity; further, as described above, a perfect ordering exists and can be found in $\mathcal{O}(m)$ time. This gives the total run-time complexity of $\mathcal{O}(nm)$. Likewise, we stated above that for a given constant $\kappa$, it can be determined in $\mathcal{O}(n)$ time whether a graph has treewidth $w^* \leq \kappa$, and if so, a vertex ordering $d$ with $w_d = w^*$ can be found within the same time bound. Then, omitting the constant factor $w_d$, the algorithm runs in $\mathcal{O}(n^2)$ time. This also follows from the algorithm's pseudocode by noting that every vertex $k$ has a constant number (at most $w^*$) of neighbours $j < k$.

We note here the similarity between Snowball and the P³C algorithm (Planken et al., 2008), presented below. Like Snowball, P³C operates by enforcing DPC, followed by a single backward sweep along the vertex ordering. P³C then computes, in $\mathcal{O}(nw_d^2)$ time, shortest

---

**Algorithm 5:** P³C (Planken et al., 2008)

---

 **Input**: Weighted directed graph $G = \langle V, E \rangle$; vertex ordering $d : V \to \{1, \ldots, n\}$
 **Output**: PPC version of $G$, or INCONSISTENT if $G$ contains a negative cycle

**1** $G \leftarrow \mathsf{DPC}(G, d)$
**2 return** INCONSISTENT **if** DPC did

**3 for** $k \leftarrow 1$ **to** $n$ **do**
**4**  $\quad$ **forall** $i, j < k$ **such that** $\{i, k\}, \{j, k\} \in E$ **do**
**5**  $\quad\quad$ $w_{i \to k} \leftarrow \min \{w_{i \to k}, w_{i \to j} + w_{j \to k}\}$
**6**  $\quad\quad$ $w_{k \to j} \leftarrow \min \{w_{k \to j}, w_{k \to i} + w_{i \to j}\}$
**7**  $\quad$ **end**
**8 end**

**9 return** $G$

---

paths only for the arcs present in the chordal graph. This similarity and a property of chordal graphs in fact prompt us to present a version of Snowball with improved time complexity.

### 3.3 Improving Run-Time Complexity Using Separators

In this section, we present an improvement of Snowball for an $\mathcal{O}\left(nw_d^2 + n^2 s_d\right)$ run time, where $s_d$ is the size of the largest minimal separator in the chordal graph obtained by triangulation along $d$.

**Definition 4.** *Given a connected graph $G = \langle V, E \rangle$, a separator is a set $V' \subseteq V$ such that $G_{V \setminus V'}$ is no longer connected. A separator $V'$ is* minimal *if no proper subset of $V'$ is a separator.*

This bound is better because, as seen below, it always holds that $s_d \leq w_d$. The improvement hinges on a property of chordal graphs called *partial path consistency* (PPC). In a partially path-consistent graph, each arc is labelled by the length of the shortest path between its endpoints.[3] P³C, presented as Algorithm 5, depends on DPC and computes PPC in $\mathcal{O}\left(nw_d^2\right)$ time, which is the current state of the art. Then, we use a *clique tree* of the PPC graph to compute the shortest path between all vertices. Figure 2 shows an example of a chordal graph and its associated clique tree. Such a clique tree has the following useful properties (Heggernes, 2006, Section 3.2).

**Property 1.** *Every chordal graph $G = \langle V, E \rangle$ has an associated clique tree $T = \langle C, S \rangle$, which can be constructed in linear time $\mathcal{O}\left(m_c\right)$.*

**Property 2.** *Each clique tree node $c \in C$ is associated with a subset $V_c \subseteq V$ and induces a maximal clique in $G$. Conversely, every maximal clique in $G$ has an associated clique tree node $c \in C$.*

**Property 3.** *$T$ is* coherent*: for each vertex $v \in V$, the clique tree nodes whose associated cliques contain $v$ induce a subtree of $T$.*

---

3. Full path-consistency (FPC) is achieved if an arc exists for all pairs of vertices $u, v \in V$.

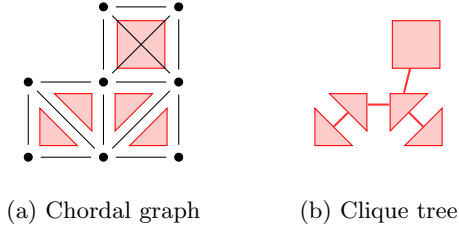

(a) Chordal graph      (b) Clique tree

Figure 2: A chordal graph and its clique tree. Each shaded shape represents a maximal clique of the graph, containing the vertices at its corners.

**Property 4.** *If two clique tree nodes $c_i, c_j \in C$ are connected by an edge $\{c_i, c_j\} \in S$, $V_{c_i} \cap V_{c_j}$ is a minimal separator in $G$. Conversely, for each minimal separator $V'$ in $G$, there is a clique tree edge $\{c_i, c_j\} \in S$ such that $V' = V_{c_i} \cap V_{c_j}$.*

**Property 5.** *All vertices appear in at least one clique associated with a node in $T$, so: $\bigcup_{c \in C} V_c = V$.*

Since we have by Proposition 2 on page 358 that the size of the largest clique in a chordal graph is exactly $w_d + 1$, it follows from Properties 2 and 4 that $s_d \leq w_d$.

Now, the idea behind Snowball–Separators is to first compute PPC in $\mathcal{O}\left(nw_d^2\right)$ time using $\mathsf{P}^3\mathsf{C}$, and then traverse the clique tree. PPC ensures that shortest paths within each clique have been computed. Then, when traversing the clique tree from an arbitrary root node out, we grow a set $V_{\mathrm{visited}}$ of vertices in cliques whose nodes have already been traversed. For each clique node $c \in C$ visited during the traversal, shortest paths between vertices in the clique $V_c$ and vertices in $V_{\mathrm{visited}}$ must run through the separator $V_{\mathrm{sep}}$ between $c$ and $c$'s parent. If $s_d$ is the size of the largest minimal separator in $G$, for each pair of vertices it suffices to consider at most $s_d$ alternative routes for a total of $\mathcal{O}\left(n^2 s_d\right)$ routes, yielding the stated overall time complexity of $\mathcal{O}\left(nw_d^2 + n^2 s_d\right)$. We formally present the algorithm based on this idea as Algorithm 6 with its associated recursive procedure Process–clique–tree–node (on the following page).

Note that because we visit a node's parent before visiting the node itself, it always holds that $V_{c_{\mathrm{parent}}} \subseteq V_{\mathrm{visited}}$. Further note that, for simplicity of presentation, we assume the graph to be connected. If not, we can simply find all connected components in linear time and construct a clique tree for each of them.

The improved algorithm has an edge over the original algorithm when separators are small while the treewidth is not. HTN-based sibling-restricted STNs (which are described as part of our experimental validation in Section 4.3.5), for instance, have many separators of size 2. If every task has as many as $\mathcal{O}\left(\sqrt{n}\right)$ subtasks and every task with its subtasks induces a clique, we have $w_d \in \mathcal{O}\left(\sqrt{n}\right)$ and $s_d = 2$, implying that Snowball–Separators still has an optimal $\mathcal{O}\left(n^2\right)$ time complexity for these instances.[4]

Before we proceed to prove that the algorithm is correct and meets the stated run-time bounds, we introduce the following definition.

---

4. However, since in general not every task and its subtasks form a clique, this low value of $s_d$ will usually not be attained in practice.

---

**Algorithm 6:** Snowball–separators

---

**Input**: Weighted directed graph $G = \langle V, E \rangle$; vertex ordering $d : V \to \{1, \ldots, n\}$

**Output**: Distance matrix $D$, or INCONSISTENT if $G$ contains a negative cycle

**1** $G \leftarrow \mathsf{P^3C}(G, d)$

**2 return** INCONSISTENT **if** $\mathsf{P^3C}$ did

**3** $\forall i, j \in V : D[i][j] \leftarrow \infty$

**4** $\forall i \in V : D[i][i] \leftarrow 0$

**5** $\forall \{i, j\} \in E : D[i][j] \leftarrow w_{i \to j}$

**6** $\forall \{i, j\} \in E : D[j][i] \leftarrow w_{j \to i}$

**7** build a clique tree $T = \langle C, S \rangle$ of $G$

**8** select an arbitrary root node $c_{\text{root}} \in C$ of $T$

**9** $(D, V_{\text{visited}}) \leftarrow$ Process–clique–tree–node$(c_{\text{root}}, \mathsf{nil}, D, \varnothing)$

**10 return** $D$

---

---

**Procedure** Process–clique–tree–node$(c, c_{\text{parent}}, D, V_{\text{visited}})$

---

**Input**: Current clique tree node $c$, $c$'s parent $c_{\text{parent}}$, distance matrix $D$, set of visited vertices $V_{\text{visited}}$

**Output**: Updated matrix $D$ and set $V_{\text{visited}}$

**1 if** $c_{\text{parent}} \neq \mathsf{nil}$ **then**

**2** $\quad$ $V_{\text{new}} \leftarrow V_c \setminus V_{c_{\text{parent}}}$

**3** $\quad$ $V_{\text{sep}} \leftarrow V_c \cap V_{c_{\text{parent}}}$

**4** $\quad$ $V_{\text{other}} \leftarrow V_{\text{visited}} \setminus V_c$

**5** $\quad$ **forall** $(i, j, k) \in V_{\text{new}} \times V_{\text{sep}} \times V_{\text{other}}$ **do**

**6** $\quad\quad$ $D[i][k] \leftarrow \min \{D[i][k], D[i][j] + D[j][k]\}$

**7** $\quad\quad$ $D[k][i] \leftarrow \min \{D[k][i], D[k][j] + D[j][i]\}$

**8** $\quad$ **end**

**9 end**

**10** $V_{\text{visited}} \leftarrow V_{\text{visited}} \cup V_c$

**11 forall** children $c'$ of $c$ **do**

**12** $\quad$ $(D, V_{\text{visited}}) \leftarrow$ Process–clique–tree–node$(c', c, D, V_{\text{visited}})$ $\quad\quad$ `// recursive call`

**13 end**

**14 return** $(D, V_{\text{visited}})$

---

**Definition 5.** *We define a distance matrix $D$ as* valid *for a set $U$ of vertices, and $(D, U)$ as a* valid pair, *if for all pairs of vertices $(i, j) \in U \times U$, $D[i][j]$ holds the shortest distance in $G$ from $i$ to $j$.*

We split the correctness proof of the algorithm into three parts: Lemmas 4 and 5 culminate in Theorem 6. The first step is to show that if Process–clique–tree–node is called with a valid pair $(D, U)$ and some clique node $c$, the procedure extends the validity to $U \cup V_c$.

**Lemma 4.** *Consider a call to procedure* Process–clique–tree–node *with, as arguments, a clique node $c$, $c$'s parent $c_{\mathrm{parent}}$, a distance matrix $D$, and the set of visited vertices $V_{\mathrm{visited}}$. If $D$ is valid for $V_{\mathrm{visited}}$ upon calling, then $D$ becomes valid for $V_c \cup V_{\mathrm{visited}}$ after running lines 1–8 of* Process–clique–tree–node.

*Proof.* First, note that by Property 2, $V_c$ induces a clique in $G$. Therefore, edges exist between each pair $(i, k)$ of vertices in $V_c$, and since the graph is PPC, $w_{i \to k}$ is labelled with the shortest distance between $i$ and $k$. Due to lines 5 and 6 of the main algorithm, $D$ also contains these shortest distances, so $D$ is valid for $V_c$.

Now, it remains to be shown that for each pair of vertices $(i, k) \in V_c \times V_{\mathrm{visited}}$ the shortest distances $D[i][k]$ and $D[k][i]$ are set correctly. We show here the case for $D[i][k]$; the other case is analogous.

The desired result follows trivially if $c_{\mathrm{parent}} = \mathsf{nil}$, since the procedure is then called with $V_{\mathrm{visited}} = \varnothing$. Otherwise, let $V_{\mathrm{new}} = V_c \setminus V_{c_{\mathrm{parent}}}$, $V_{\mathrm{sep}} = V_c \cap V_{c_{\mathrm{parent}}}$ and $V_{\mathrm{other}} = V_{\mathrm{visited}} \setminus V_c$ as set by the procedure in lines 2–4. If either $i$ or $k$ lies in $V_{\mathrm{sep}}$, the correctness of $D[i][k]$'s value was already proven, so we only need to consider pairs of vertices $(i, k) \in V_{\mathrm{new}} \times V_{\mathrm{other}}$.

For any such pair $(i, k)$, $V_{\mathrm{sep}}$ is a separator between $i$ and $k$ by Property 4, so any shortest path from $i$ to $k$ is necessarily a concatenation of shortest paths from $i$ to $j^*$ and from $j^*$ to $k$, for some $j^* \in V_{\mathrm{sep}}$. Since it follows from the definitions of $V_{\mathrm{new}}$, $V_{\mathrm{sep}}$ and $V_{\mathrm{other}}$ that for all $(i, j) \in V_{\mathrm{new}} \times V_{\mathrm{sep}}$ and $(j, k) \in V_{\mathrm{sep}} \times V_{\mathrm{other}}$, $D[i][j]$ and $D[j][k]$ are correctly set (by the validity of $D$ for $V_c$ and $V_{\mathrm{visited}}$, respectively), the loop on lines 5–8 yields the desired result. □

Our next step is to prove that through the recursive calls, validity is in fact extended to the entire subtree rooted at $c$.

**Lemma 5.** *Consider again a call to procedure* Process–clique–tree–node *with, as arguments, a clique node $c$, $c$'s parent $V_{c_{\mathrm{parent}}}$, a distance matrix $D$, and the set of visited vertices $V_{\mathrm{visited}}$. If $D$ is valid for $V_{\mathrm{visited}}$ upon calling, then the returned, updated pair $(D', V'_{\mathrm{visited}})$ is also valid.*

*Proof.* First, note that by Lemma 4, $D$ is valid for $V_{\mathrm{visited}}$ after the update in line 10.

Assume that the clique tree has a depth of $d$; the proof is by reverse induction over the depth of the clique tree node. If $c$ is a clique tree node at depth $d$ (i.e. a leaf), the loop in lines 11–13 is a no-op, so we immediately obtain the desired result.

Now assume that the lemma holds for all nodes at depth $k$ and let $c$ be a clique tree node at depth $k - 1$. For the first call (if any) made for a child node $c'$ during the loop in lines 11–13, this lemma can then be applied. As a consequence, the returned and updated pair is again valid. This argument can be repeated until the loop ends and the procedure returns a valid pair. □

With these results at our disposal, we can state and prove the main theorem of this section.

**Theorem 6.** *Algorithm 6 (Snowball–Separators) correctly computes all-pairs shortest paths in $\mathcal{O}\left(nw_d^2 + n^2 s_d\right)$ time.*

*Proof.* Note that $V_\text{visited} = \varnothing$ for the call to Process–clique–tree–node in line 9 of Snowball–Separators; therefore, the pair $(D, V_\text{visited})$ is trivially valid. By Lemma 5, this call thus returns a valid updated pair $(D, V_\text{visited})$. Since Process–clique–tree–node has recursively traversed the entire clique tree, $V_\text{visited}$ contains the union $\bigcup_{c \in C} V_c$ of all cliques in the clique tree $T = \langle C, S \rangle$, which by Property 5 equals the set of all vertices in $G$. Therefore, $D$ contains the correct shortest paths between all pairs of vertices in the graph.

As for the time complexity, note that the initialisations in lines 3 and 4 can be carried out in $\mathcal{O}\left(n^2\right)$ time, whereas those in lines 5 and 6 require $\mathcal{O}\left(m_c\right)$ time. By Property 1, the clique tree can be built in linear time $\mathcal{O}\left(m_c\right)$. Since the clique tree contains at most $n$ nodes, Process–clique–tree–node is called $\mathcal{O}\left(n\right)$ times. Line 1 requires $\mathcal{O}\left(w_d^2\right)$ time. To implement lines 2–4 and 10 of Process–clique–tree–node, we represent the characteristic function for $V_\text{visited}$ as an array of size $n$; using $V_\text{visited}$ instead of $V_{c_\text{parent}}$ everywhere, we then we simply iterate over all $\mathcal{O}\left(w_d\right)$ members of $V_c$ to perform the required computations.

Now, only the complexity of the loop in lines 5–8 remains to be shown. Note that $|V_\text{sep}| \leq s_d$ by definition, and $|V_\text{other}| < n$ always. Further using the observation that each of the $n$ vertices in the graph appears in $V_\text{new}$ for exactly one invocation of Process–clique–tree–node (after which it becomes a staunch member of $V_\text{visited}$), we obtain a total time bound of $\mathcal{O}\left(n^2 s_d\right)$ for the loop over all invocations. □

While the recursive description above is perhaps easier to grasp and satisfies the claimed time bounds, we found that efficiency benefited in practice from an iterative implementation. It also turns out that a good heuristic is to first visit child nodes connected to the already visited subtree by a large separator, postponing the processing of children connected by a small separator, because the set of visited vertices is then still small. In this way, the sum of terms $|V_\text{sep} \times V_\text{visited}|$ is kept low. In our implementation, we therefore use a priority queue of clique nodes ordered by their separator sizes. Future research must point out whether it is feasible to determine an optimal traversal of the clique tree within the given time bounds.

Having presented our new algorithms and proven their correctness and formal complexity, we now move on to an empirical evaluation of their performance.

## 4. Experiments

We evaluate the two algorithms together with efficient implementations of Floyd–Warshall and Johnson with a Fibonacci heap[5] across six different benchmark sets.[6]

---

Table 1: Properties of the benchmark sets

| type | #cases | $n$ | $m$ | $w_d$ | $s_d$ |
|---|---|---|---|---|---|
| Chordal | | | | | |
| – Figure 3 | 250 | 1,000 | 75,840–499,490 | 79–995 | 79–995 |
| – Figure 4 | 130 | 214–3,125 | 22,788–637,009 | 211 | 211 |
| Scale-free | | | | | |
| – Figure 5 | 130 | 1,000 | 1,996–67,360 | 88–864 | 80–854 |
| – Figure 6 | 160 | 250–1,000 | 2,176–3,330 | 150–200 | 138–190 |
| New York | 170 | 108–3,906 | 113–6,422 | 2–51 | 2–40 |
| Diamonds | 130 | 111–2,751 | 111–2,751 | 2 | 2 |
| Job-shop | 400 | 17–1,321 | 32–110,220 | 3–331 | 3–311 |
| HTN | 121 | 500–625 | 748–1,599 | 2–128 | 2–127 |

The properties of the test cases are summarised in Table 1. This table lists the number of test cases, the range of the number of vertices $n$, edges $m$, the induced width $w_d$ produced by the minimum degree heuristic, as well as the size of the largest minimal separators $s_d$ in the graphs. More details on the different sets can be found below, but one thing that stands out immediately is that $s_d$ is often equal to or only marginally smaller than $w_d$. However, the median size of the minimum separator is less than 10 for all instances except the constructed chordal graphs.

All algorithms were implemented in Java and went through an intensive profiling phase.[7] The experiments were run using Java 1.6 (OpenJDK-1.6.0.b09) in server mode, on Intel Xeon E5430 CPUs running 64-bit Linux. The Java processes were allowed a maximum heap size of 4 GB, and used the default stack size. We report the measured CPU times, including the time that was spent running the triangulation heuristic for Chleq–APSP and Snowball. The reported run times are averaged over 10 runs for each unique problem instance. Moreover, we generated 10 unique instances for each parameter setting, obtained by using different random seeds. Thus, each reported statistic represents an average over 100 runs, unless otherwise indicated. Finally, each graph instance was ensured to contain no cycles of negative weight.

## 4.1 Triangulation

As discussed in Section 2.2, finding an optimal vertex ordering (with minimum induced width) is NP-hard, but several efficient triangulation heuristics for this problem exist. We ran our experiments with six different heuristics: the minimum fill and minimum degree heuristics, static variants of both (taking into account only the original graph), an ordering produced by running maximum cardinality search (MCS) on the original graph, and a random ordering. All of these, except minimum fill, have time complexities within the bound on the run time of Chleq–APSP and Snowball. We found that the minimum degree heuristic gave on average induced widths less than 1.5% higher than those found by minimum fill, but with drastically lower run time. The exorbitant time consumption of the minimum fill heuristic can be partially explained by the fact that we used the LibTW package[8] to compute this ordering, whose implementation can probably be improved. However, it is also known from the literature that the theoretical bound on the minimum fill heuristic is worse than that of minimum degree (Kjærulff, 1990). All other heuristics are not only slower than minimum degree, but also yield an induced width at least 12% higher, resulting in a longer total triangulation time and a longer total run time of Snowball (see the summary of the results over all benchmarks given in Table 2). Again, this confirms Kjærulff's earlier work. In the experimental results included below we therefore only show the results based on the minimum degree heuristic.

---

7. Our implementations are available in binary form at
   http://dx.doi.org/10.4121/uuid:776a266e-81c6-41ee-9d23-8c89d90b6992

Table 2: The summed induced width, triangulation, and total run time of Snowball over all experiments on general (non-chordal) graphs show that the minimum degree heuristic is the best choice.

| heuristic | $\sum w_d$ | triangulation (s) | Snowball (s) | total (s) |
|---|---|---|---|---|
| min-fill | 321,492 | 1,204,384 | 2,047 | 1,206,431 |
| min-degree | 326,222 | 498 | 3,166 | 3,664 |
| MCS | 365,662 | 1,520 | 3,348 | 4,868 |
| static min-fill | 388,569 | 1,387 | 2,746 | 4,133 |
| static min-degree | 388,707 | 1,317 | 2,748 | 4,064 |
| random | 505,844 | 2,436 | 5,179 | 7,615 |

## 4.2 Chordal Graphs

To evaluate the performance of the new algorithms on chordal graphs, we construct chordal graphs of a fixed size of 1,000 vertices with a treewidth ranging from 79 up to just less than the number of vertices, thus yielding a nearly complete graph at the high end. The results of this experiment are depicted in Figure 3. In this, and other figures, the error bars represent the standard deviations in the measured run time for the instances of that size. For graphs up to an induced width of about three quarters of the number of vertices, Snowball significantly outperforms Floyd–Warshall (which yields the expected horizontal line), and overall the run time of both new algorithms is well below that of Johnson across the entire range. Figure 4 shows the run times on chordal graphs of a constant treewidth and with increasing number of vertices. Here, the two new algorithms outperform Johnson by nearly an order of magnitude (a factor 9.3 for Snowball around $n = 1300$), and even more so regarding Floyd–Warshall, confirming the expectations based on the theoretical upper bounds.

## 4.3 General Graphs

For general, non-chordal graphs, we expect from the theoretical analysis that the $\mathcal{O}\left(nw_d^2\right)$-time Chleq–APSP and Snowball algorithms are faster than Johnson with its $\mathcal{O}\left(nm + n^2 \log n\right)$ time bound when $w_d$ is low, and that Johnson is faster on sparse graphs (where $m$ is low) of a large induced width $w_d$. The main question is at which induced width this changeover occurs. Regarding Floyd–Warshall with its $\mathcal{O}\left(n^3\right)$ bound, we expect that for larger $n$ it is always outperformed by the other algorithms.

---

8. Available from http://treewidth.com/.

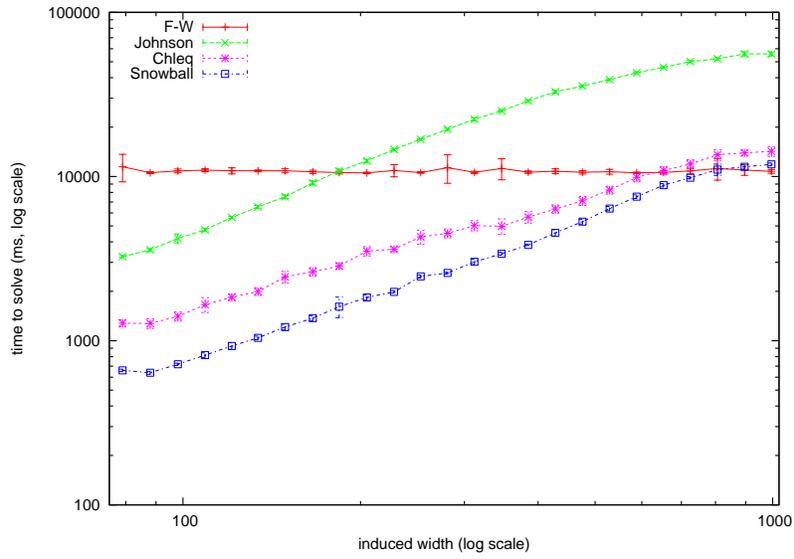

Figure 3: Run times on generated chordal graphs with a fixed number of 1000 vertices and varying treewidth.

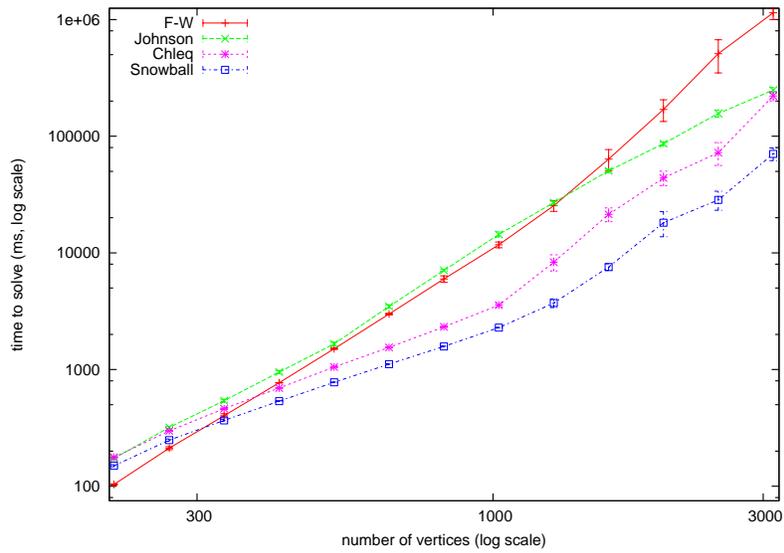

Figure 4: Run times on generated chordal graphs of a fixed treewidth of 211.

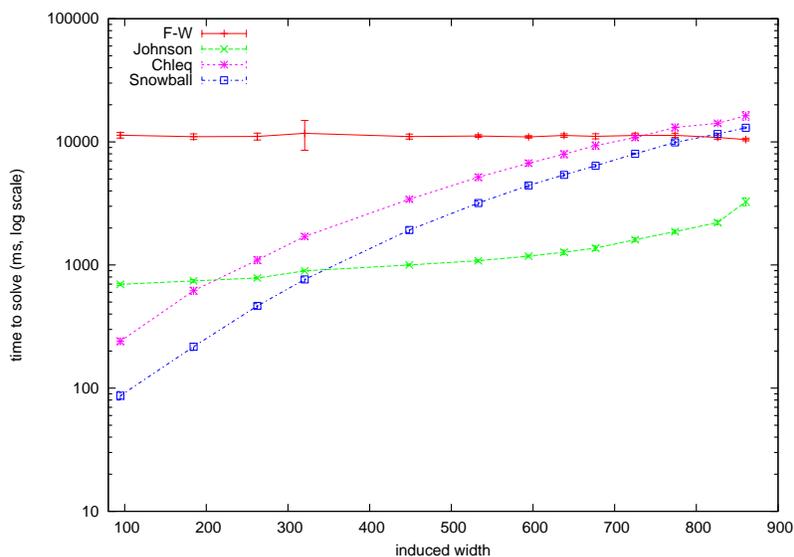

Figure 5: Run times on the scale-free benchmarks for graphs of 1,000 vertices and varying induced width.

### 4.3.1 Scale-Free Graphs

Scale-free networks are networks whose degree distribution follows a power law. That is, for large values of $k$, the fraction $P(k)$ of vertices in the network having $k$ connections to other vertices tends to $P(k) \sim ck^{-\gamma}$, for some constant $c$ and parameter $\gamma$. In other words, few vertices have many connections while many vertices have only a few connections. Such a property can be found in many real-world graphs, such as in social networks and in the Internet. Our instances were randomly generated with Albert and Barabási's (2002) preferential attachment method, where in each iteration a new vertex is added to the graph, which is then attached to a number of existing vertices; the higher the degree of an existing vertex, the more likely it is that it will be connected to the newly added vertex. To see at which induced width Johnson is faster, we compare the run times on such generated graphs with 1,000 vertices. By varying the number of attachments for each new vertex from 2 to $n/2$, we obtain graphs with an induced width ranging from 88 to 866. In these graphs, the induced width is already quite large for small attachment values: for example, for a value of 11, the induced width is already over 500.

The results of this experiment can be found in Figure 5. Here we see that up to an induced width of about 350 (attachment value 5), Snowball is the most efficient. For higher induced widths, Johnson becomes the most efficient; for $w_d$ around 800, even Floyd–Warshall becomes faster than Snowball. A consistent observation but from a different angle can be made from Figure 6, where the induced width is between 150 and 200, the number of edges is between 2,176 and 3,330 and the number of vertices is varied from 250 to 1,000. Here we see that for small graphs up to 350 vertices, Johnson is the fastest; then Snowball overtakes it, and around 750 vertices Chleq-APSP is also faster than Johnson (this holds for all results up to a sparse graph of 1,000 vertices).

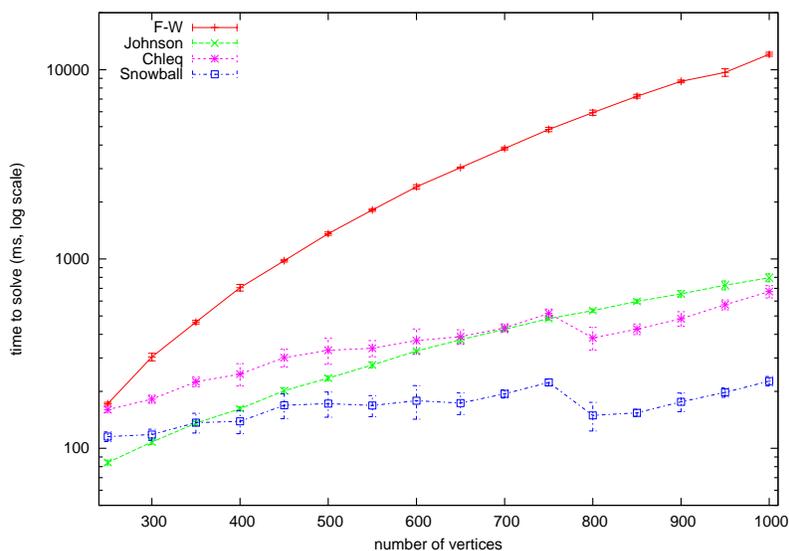

Figure 6: Run times on the scale-free benchmarks for graphs of induced widths 150 to 200 and varying vertex count.

Around the mark of 750 vertices, the results show a decrease in the run time for both Snowball and Chleq-APSP. This is an artifact of the (preferential attachment) benchmark generator. Since we cannot generate scale-free graphs with a specific induced width, we modify the attachment value instead. As it turns out, for graphs of this size only one attachment value yields an induced width within the desired range; for the graph of size 750, this width is at the high end of the interval, whereas for the graph of size 800 it is near the low end. This explains the reduced run time for the larger graph.

For these scale-free networks, we conclude that Snowball is the fastest of the four algorithms when the induced width is not too large (at most one third of the number of vertices in our benchmark set). However, we also observe that the structure of scale-free networks is such that they have a particularly high induced width for relatively sparse graphs, exactly because a few vertices have most of the connections. Therefore, Snowball is most efficient only for relatively small attachment values.

### 4.3.2 Selections from New York Road Network

More interesting than the artificially constructed graphs are graphs based on real networks, for which shortest path calculations are relevant. The first of this series is based on the road network of New York City, which we obtained from the DIMACS challenge website.[9] This network is very large (with 264,346 vertices and 733,846 edges) so we decided to compute shortest paths for (induced) subgraphs of varying sizes. These were obtained by running a simple breadth-first search from a random starting location until the desired number of vertices had been visited. The extent of the subnetworks thus obtained is illustrated for three different sizes in Figure 7. The results of all algorithms on these subgraphs can be found in Figure 8. Here we observe the same ranking of the algorithms as on the chordal graphs of a fixed treewidth and for diamonds: Floyd–Warshall is slowest with its $\Theta\left(n^3\right)$ run time, then each of Johnson, Chleq–APSP, and Snowball is significantly faster than its predecessor. This can be explained by considering the induced width of these graphs. Even for the largest graphs the induced width is around 30, which is considerably smaller than the number of vertices.

---

9. http://www.dis.uniroma1.it/~challenge9/

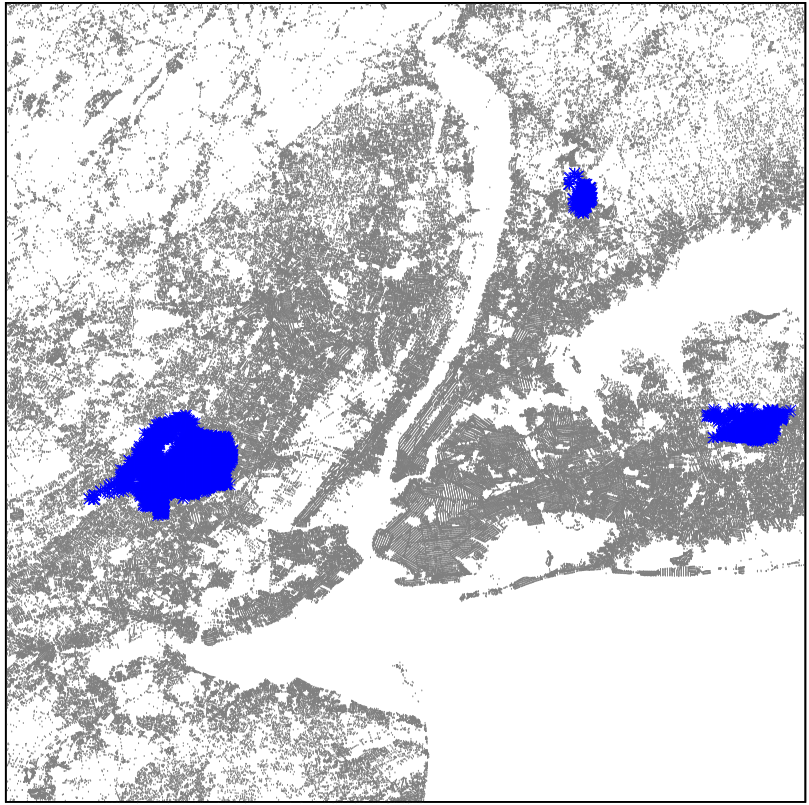

Figure 7: Coordinates for the vertices in the New York City input graph, and examples of the extent of subgraphs with respectively 250, 1000, and 5000 vertices.

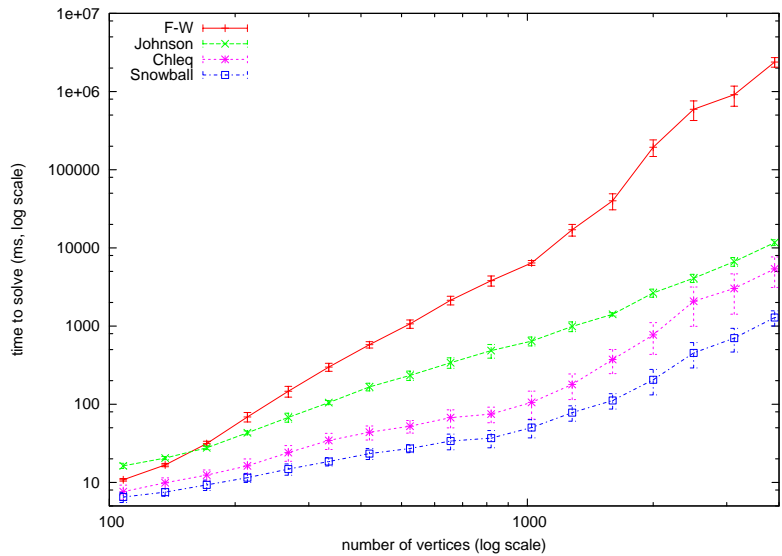

Figure 8: Run times on the New York benchmarks for subgraphs of varying vertex count.

### 4.3.3 STNs from Diamonds

This benchmark set is based on problem instances in difference logic proposed by Strichman, Seshia, and Bryant (2002) and also appearing in the smt-lib (Ranise & Tinelli, 2003), where the constraint graph for each instance takes the form of a circular chain of "diamonds". Each such diamond consists of two parallel paths of equal length starting from a single vertex and ending in another single vertex. From the latter vertex, two paths start again, to converge on a third vertex. This pattern is repeated for each diamond in the chain; the final vertex is then connected to the very first one. The sizes of each diamond and the total number of diamonds are varied between benchmarks.

Problems in this class are actually instances of the NP-complete Disjunctive Temporal Problem (DTP): constraints take the form of a disjunction of inequalities. From each DTP instance, we obtain a STP instance (i.e. a graph) by randomly selecting one inequality from each such disjunction. This STP is most probably inconsistent, so its constraint graph contains a negative cycle; we remedy this by modifying the weights on the constraint edges. The idea behind this procedure is that the *structure* of the graph still conforms to the type of networks that one might encounter when solving the corresponding DTP instance, and that the run time of the algorithms mostly depends on this structure. Moreover, to reduce the influence of the randomized extraction procedure, we repeat it for 10 different seeds.

For our benchmark set, we considered problem instances which had the size of the diamonds fixed at 5 and their number varying. The most interesting property of this set is that the graphs generated from it are very sparse. We ran experiments on 130 graphs, ranging in size from 111 to 2751 vertices, all with an induced width of 2. This induced width is clearly extremely small, which translates into Chleq–APSP and Snowball being considerably faster than Johnson and Floyd–Warshall, as evidenced by Figure 9.

### 4.3.4 STNs from Job-Shop Scheduling

We generated each of the 400 graphs in our "job-shop" set from an instance of a real job-shop problem. These instances were of the type available in smt-lib (Ranise & Tinelli, 2003), but of a larger range than included in that benchmark collection. To obtain these graphs from the job-shop instances, we again used the extraction procedure described in the previous section. The most striking observation that can be taken from Figure 10 is that the difference between Johnson and the two new algorithms is not quite as pronounced, though Snowball is consistently the fastest of the three by a small margin. The fact that this margin is so small is most likely due to the structure of these graphs, which is also reflected in their relatively high induced width. Note also that the run times for Floyd–Warshall are better for graphs of up to 160 vertices, while for larger graphs the other algorithms are significantly faster.

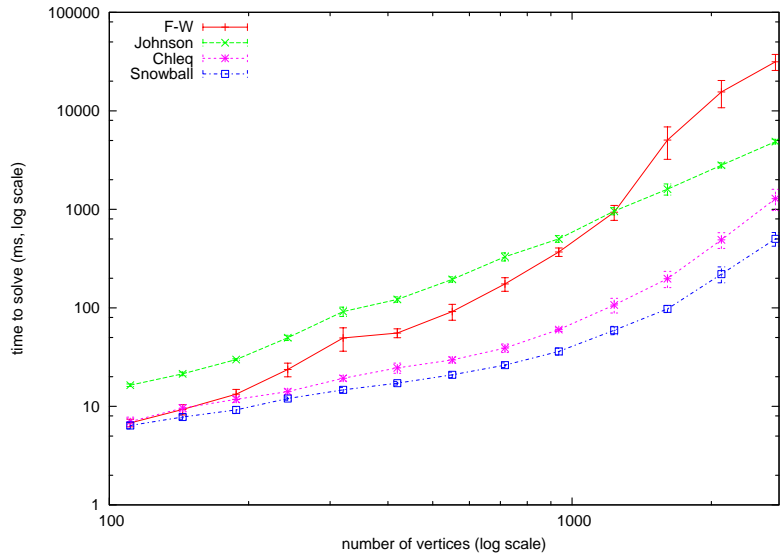

Figure 9:  Run times on the diamonds benchmarks for graphs of varying vertex count.

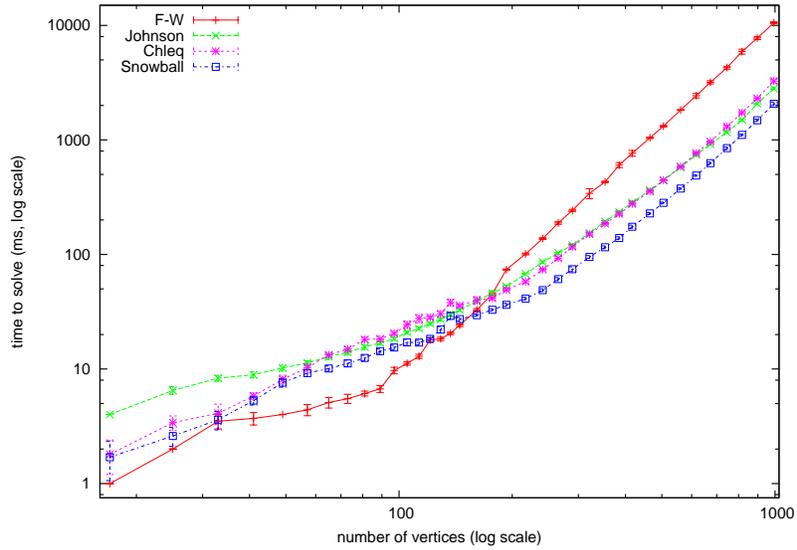

Figure 10:  Run times on the job-shop benchmarks for graphs of varying vertex count.

### 4.3.5 STNs from HTNs

Finally, we consider a benchmark set whose instances imitate so-called sibling-restricted STNs originating from Hierarchical Task Networks. This set is therefore particularly interesting from a planning point of view. In these graphs, constraints may occur only between parent tasks and their children, and between sibling tasks (Bui & Yorke-Smith, 2010). We consider an extension that includes *landmark variables* (Castillo, Fernández-Olivares, & González, 2002) that mimic synchronisation between tasks in different parts of the network, and thereby cause some deviation from the tree-like HTN structure. We generate HTNs using the following parameters: (i) the number of tasks in the initial HTN tree (fixed at 250; note that tasks have a start and end point), (ii) the branching factor, determining the number of children for each task (between 4 and 7), (iii) the depth of the HTN tree (between 3 and 7), (iv) the ratio of landmark time points to the number of tasks in the HTN, varying from 0 to 0.5 with a step size of 0.05, and (v) the probability of constraints between siblings, varying from 0 to 0.5 with a step size of 0.05.

These settings result in graphs of between 500 and 625 vertices, with induced widths varying between 2 and 128. Though the induced width seems high in light of our claim above that it is constant, we verified that $w_d \leq 2 \times branching\ factor + \#landmarks + 1$ for all instances. Filling in the maximal values of 7 and 125 respectively, we find an upper bound $w_d \leq 140$, well above the actual maximum encountered.

Figure 11 shows the results of these experiments as a function of the induced widths of the graphs. We can see that only for the larger induced widths, Johnson and Chleq–APSP come close. These large induced widths are only found for high landmark ratios of 0.5. The results indicate that for the majority of STNs stemming from HTNs, Snowball is significantly more efficient than Johnson.

### 4.4 Snowball–Separators

In Section 3.3 we presented a version of Snowball that has an improved worst-case run time over vanilla Snowball by taking advantage of the separators in the graph. In this section, we discuss the results of our experiments comparing these two variants. First, we turn our attention to the benchmark problems on regular graphs. Our results are summarised in Figure 12. As one can see, Snowball–Separators actually performs strictly worse on these sets in terms of run-time performance when compared to the original Snowball.

However, as can be seen in Table 1, the largest minimal separator is often equal to or only marginally smaller than the induced width. Even though there may be only few separators this large, and many may be substantially smaller (as noted above, for most instances the median separator size is below 10), this prompts us to run experiments on instances where separator sizes are artificially kept small. Indeed, we found that there are cases where Snowball–Separators shows an improvement over vanilla Snowball when comparing the number of update operations performed—i.e. lines 8 and 9 of Snowball and lines 6 and 7 in Process–clique–tree–node, along with lines 3 and 4 of DPC and lines 5 and 6 of P³C. One such case is presented in Figure 13. This describes the results on a collection of chordal graphs of 512 vertices, in which the largest minimal separator is fixed at size 2, and the treewidth is varied between 16 and 448. The figure also includes the results of DPC and P³C, as these are the respective subroutines of Snowball and Snowball–Separators. For these graphs, Snowball–Separators performs strictly fewer update operations than Snowball on all instances, although the difference becomes smaller as the induced width increases. While the number of updates shows a distinct improvement over Snowball, the run times of the Snowball–Separators algorithms do not show the same improvement. Instead, as can be seen from Figure 14, the run times of Snowball are strictly better than those of Snowball–Separators on all instances. Snowball can now even be seen to outperform P³C which has a better theoretical bound; the reason is that the adjacency matrix data structure as used by Snowball is very fast, while the adjacency list used by P³C, though staying within the theoretical bound, inflicts a larger constant factor on the run time.

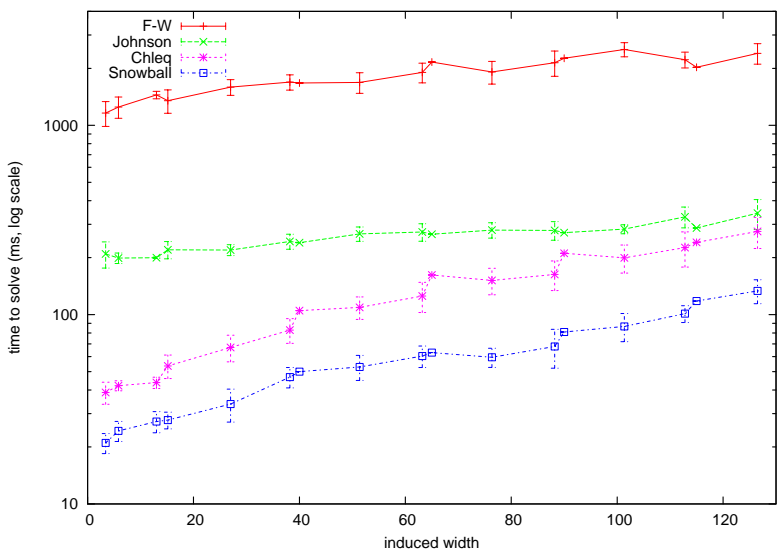

Figure 11: Run times on the HTN benchmarks for graphs from 500 to 625 vertices and varying induced width. Each point is the average of instances with an induced width within a range $[5k, 5k+4]$, for some $k$. This results in between 5 and 11 instances per data point.

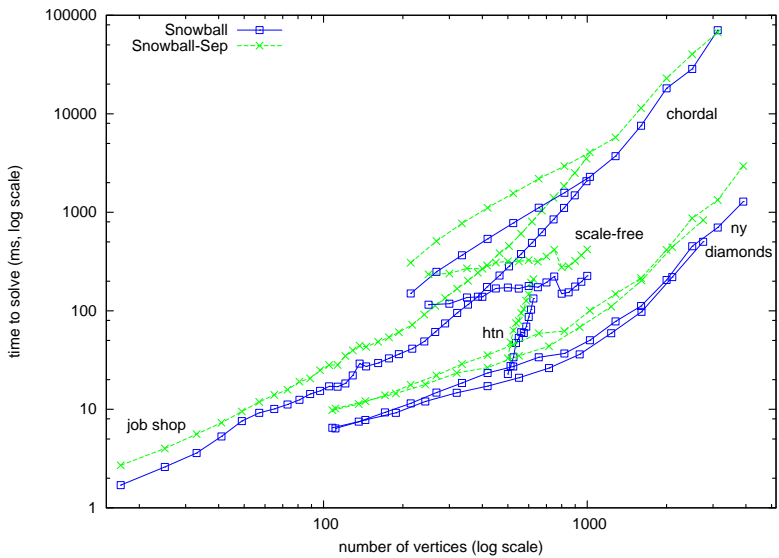

Figure 12: Run times of the Snowball algorithms on the benchmark problem sets listed in Table 1.

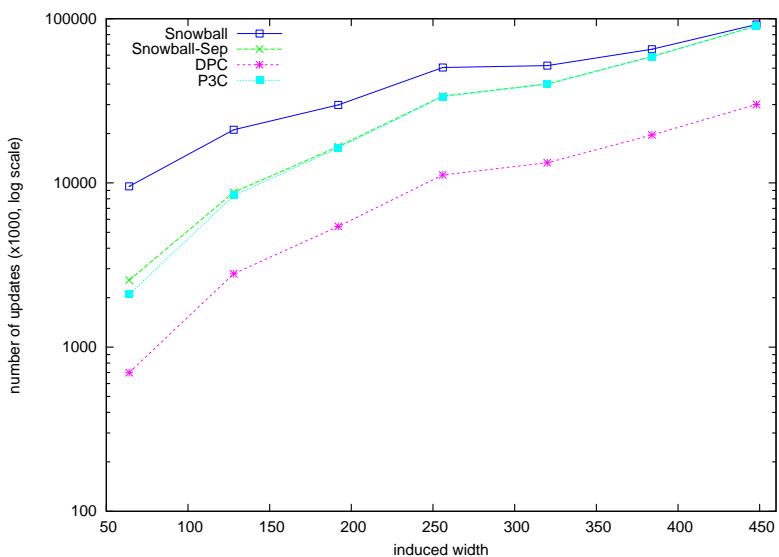

Figure 13: Number of distance matrix updates on chordal instances with 512 vertices, largest minimal separator size 2 and varying treewidth. Each point represents between 5 and 10 instances.

From these experiments, we can conclude that on graphs of these sizes, the additional bookkeeping required by Snowball–Separators outweighs the potential improvement in the number of distance matrix updates.

## 4.5 A Proper Upper Bound on the Run Time

On general graphs, the run time of the proposed algorithms depends on the induced width $w_d$ of the ordering produced by the triangulation heuristic. This induced width is not a direct measure of the input (graph), so the given upper bound on the run time is not quite proper. To arrive at a proper bound, in this section we aim to relate the run time to the treewidth, denoted $w^*$, which is a property of the input. However, determining the treewidth, an NP-hard problem, is an intractable task for the benchmark problems we used. We therefore compare the measured induced width $w_d \geq w^*$, an upper bound on the treewidth, to a lower bound $x \leq w^*$.[10] We are unaware of any guarantee on the quality relative to the treewidth of either the minimum degree triangulation heuristic or the lower bound we used. However, we can calculate the ratio $w_d/x$ to get an upper bound on the ratio $w_d/w^*$. From this measure we can then obtain an upper bound on the run time expressed in the treewidth, at least for the benchmark problems in this paper.

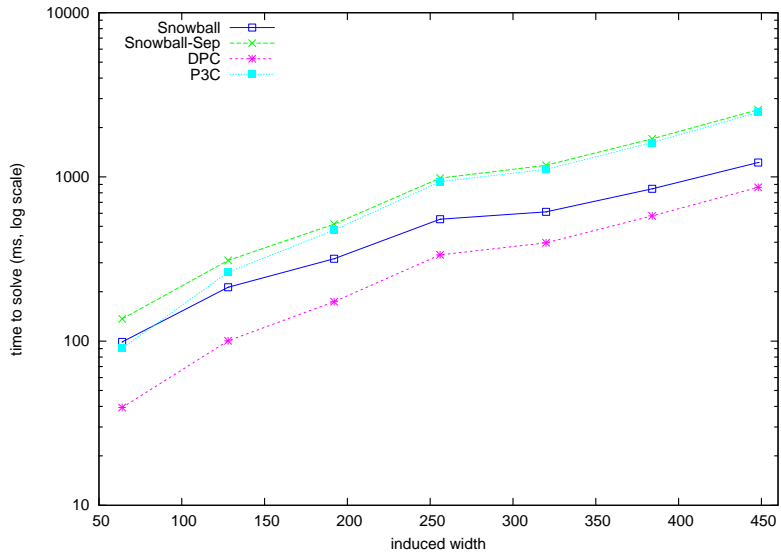

Figure 14: Run times on chordal instances with 512 vertices, largest minimal separator size 2 and varying treewidth. Each point represents between 5 and 10 instances.

The results of these computations can be found in Figure 15, where we plot these ratios for the New York, HTN, scale-free and job-shop benchmarks as a function of the lower bound $x$. Using a least-squares approach, we then fitted functions $w_d(x) = cx^k$ (showing up as a straight line in this log-log plot) to the plotted data points. For functions found by fitting, we get $k = 4.6$ for New York, $k = 2.3$ for HTN, and $k = 0.98$ for job-shop, all with small multiplicative constants $0.012 < c < 1.62$. As one can see from the plotted data points for the scale-free instances, they are not amenable to such a fit and we therefore omit it from the figure.

The decreasing trend for the job-shop data indicates that the quality of the triangulation (i.e. of the upper bound represented by the induced width) gradually increases: the lower and upper bound are always less than a factor 2 apart. Indeed, if we plot a line representing a function $w'_d(x) = 2x$ (yielding a horizontal line in this figure), we find that it describes a comfortable upper bound on the data points for this benchmark set.

The HTN data prompts us to plot a function $w''_d(x) = \frac{2}{5}x^{2.5}$, with an exponent slightly higher than the one we found from the least-squares fit, and further tweaked slightly by a multiplicative coefficient to bring it into view. This function as plotted represents an ample upper bound for the HTN benchmarks (as well as the job-shop ones).

---

10. The lower bound was computed with the LibTW package; see `http://treewidth.com/`. We used the MMD + Least-c heuristic.

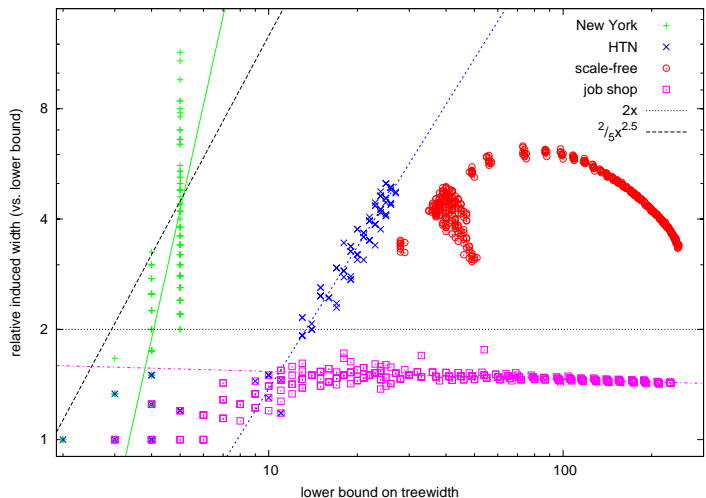

Figure 15: An upper bound on the induced width relative to the treewidth can be determined experimentally by comparing it to a lower bound on the treewidth.

The fit for the data points for the New York benchmark is not good and the trend of the points themselves is not very clear, because the lower bound only spans an interval from 1 to 4. Therefore, we cannot give an upper bound for this set of benchmarks with any acceptable level of confidence.

However, the scale-free data points we plotted, which could not be fitted with a function yielding a straight line, do mostly follow a clear curving trend. A hypothesis for this behaviour is that the quality of the upper and lower bound deteriorates mostly for the middle sizes of the benchmarks; smaller and larger scale-free graphs are easier to triangulate well.[11] To give an upper bound, we could plot any line on the outer hull of these data points; e.g. the horizontal line represented by $w_d(x) = 8x$ would work. The most pessimistic assumption would be to choose a function with the highest slope, and we find that the upper bound $w_d''(x) = \frac{2}{5}x^{2.5}$, found for the HTN benchmarks, also works here.

From this discussion, we may conclude that for all benchmarks we ran except for New York, $w_d(x)$ is $\mathcal{O}\left(x^{2.5}\right)$ which in turn is $\mathcal{O}\left(w^{*2.5}\right)$; the run time of the algorithms Snowball and Chleq–APSP on these instances can therefore be bounded by $\mathcal{O}\left(n^2 w^{*2.5}\right)$.

To conclude this section, we remark that an alternative to a triangulation heuristic would be to use an approximation algorithm with a bound on the induced width that can be theoretically determined. For example, Bouchitté, Kratsch, Müller, and Todinca (2004) give a $\mathcal{O}\left(\log w^*\right)$ approximation of the treewidth $w^*$. Using such an approximation would give an upper bound on the run time of Snowball of $\mathcal{O}\left(n^2 w^* \log w^*\right)$. However, the run time of obtaining this approximate induced width is $\mathcal{O}\left(n^3 \log^4 n w^{*5} \log w^*\right)$ and has a high constant as well, so their work is—for now—mainly of theoretical value.

---

11. This mirrors earlier observations by the authors.

## 5. Related Work

For dense, directed graphs with real weights, the state-of-the-art APSP algorithms run in $\mathcal{O}\left(n^3 / \log n\right)$ time (Chan, 2005; Han, 2008). These represent a serious improvement over the $\mathcal{O}\left(n^3\right)$ bound on Floyd–Warshall but do not profit from the fact that in most graphs that occur in practice, the number of edges $m$ is significantly lower than $n^2$.

This profit is exactly what algorithms for sparse graphs aim to achieve. Recently, an improvement was published over the $\mathcal{O}\left(nm + n^2 \log n\right)$ algorithm based on Johnson's (1977) and Fredman and Tarjan's (1987) work: an algorithm for sparse directed graphs running in $\mathcal{O}\left(nm + n^2 \log \log n\right)$ time (Pettie, 2004). In theory, this algorithm is thus faster than Johnson (in worst cases, for large graphs) when $m \in o\left(n \log n\right)$.[12] However, currently no implementation exists (as confirmed through personal communication with Pettie, June 2011). The upper bound of $\mathcal{O}\left(n^2 w_d\right)$ on the run time of Snowball is smaller than this established upper bound when the induced width is small (i.e. when $w_d \in o\left(\log \log n\right)$), and, of course, for chordal graphs and graphs of constant treewidth.

We are familiar with one earlier work to compute shortest paths by leveraging low treewidth. Chaudhuri and Zaroliagis (2000) present an algorithm for answering (point-to-point) shortest path queries with $\mathcal{O}\left(w_d^3 n \log n\right)$ preprocessing time and query time $\mathcal{O}\left(w_d^3\right)$. A direct extension of their results to APSP would imply a run time of $\mathcal{O}\left(n^2 w_d^3\right)$ on general graphs and $\mathcal{O}\left(nm w_d^2\right)$ on chordal graphs. Our result of computing APSP on general graphs in $\mathcal{O}\left(n^2 w_d\right)$ and in $\mathcal{O}\left(nm\right)$ on chordal graphs is thus a strict improvement.

A large part of the state-of-the-art in point-to-point shortest paths is focused on road networks (with positive edge weights). These studies have a strong focus on heuristics, ranging from goal-directed search and bi-directional search to using or creating some hierarchical structure, see for example (Geisberger, Sanders, Schultes, & Delling, 2008; Bauer, Delling, Sanders, Schieferdecker, Schultes, & Wagner, 2008). One of these hierarchical heuristics has some similarities to the idea of using chordal graphs. This heuristic is called *contraction*. The idea there is to distinguish important (core) vertices, which may be possible end points, from vertices that are never used as a start or end point. These latter vertices are then removed (bypassed) one-by-one, connecting their neighbours directly.

Other restrictions on the input graphs for which shortest paths are computed can also be assumed, and sometimes lead to algorithms with tighter bounds. For example, for *unweighted* chordal graphs, APSP lengths can be determined in $\mathcal{O}\left(n^2\right)$ time (Balachandran & Rangan, 1996; Han, Sekharan, & Sridhar, 1997) if all pairs at distance two are known. See (Dragan, 2005) for an overview and unification of such approaches. Considering only planar graphs, recent work shows that APSP be found in $\mathcal{O}\left(n^2 \log^2 n\right)$ (Klein, Mozes, & Weimann, 2010), which is an improvement over Johnson in cases where $m \in \omega\left(n \log^2 n\right)$.

In the context of planning and scheduling, a number of similar APSP problems need to be computed sequentially, potentially allowing for a more efficient approach using dynamic algorithms. Even and Gazit (1985) provide a method where addition of a single edge can require $\mathcal{O}\left(n^2\right)$ steps, and deletion $\mathcal{O}\left(n^4/m\right)$ *on average*. Thorup (2004) and Deme-trescu and Italiano (2006) later give an alternative approach with an amortized run time of $\mathcal{O}\left(n^2(\log n + \log^2 \frac{n+m}{n})\right)$. Especially in the context of planning and scheduling, it is not essential that the shortest paths between *all* time points be maintained. Often, it is sufficient when the shortest paths of a selection of pairs are maintained. Above, we already mentioned the P³C algorithm by Planken et al. (2008) for the single-shot case; Planken et al. (2010) describe an algorithm that incrementally maintains the property of partial path consistency on chordal graphs in time linear in the number of edges.

---

12. We explain our use of the notation $x \in o\left(f(n)\right)$ in Footnote 1 on page 354.

## 6. Conclusions and Future Work

In this paper we give three algorithms for computing all-pairs shortest paths, with a run time bounded by (i) $\mathcal{O}\left(n^2\right)$ for graphs of constant treewidth, matching earlier results that also required $\mathcal{O}\left(n^2\right)$ (Chaudhuri & Zaroliagis, 2000); (ii) $\mathcal{O}\left(nm\right)$ on chordal graphs, improving over the earlier $\mathcal{O}\left(nmw_d^2\right)$; and (iii) $\mathcal{O}\left(n^2w_d\right)$ on general graphs, showing again an improvement over previously known tightest bound of $\mathcal{O}\left(n^2w_d^3\right)$. In these bounds, $w_d$ is the induced width of the ordering used; experimentally we have determined this to be bounded by the treewidth to the power 2.5 for most of our benchmarks.

These contributions are obtained by applying directed path consistency combined with known graph-theoretic techniques, such as a vertex elimination and tree decomposition, to computing shortest paths. This supports the general idea that such techniques may help in solving graphically-representable combinatorial problems, but the main contribution of this article is more narrow, focusing on improving the state of the art for this single, but important problem of computing APSP.

From the results of our extensive experiments we can make recommendations as to which algorithm is best suited for which type of problems. Only for very small instances, Floyd–Warshall should be used; this is probably mostly thanks to its simplicity, yielding a very straightforward implementation with low overhead. Snowball can exploit the fact that a perfect elimination ordering can be efficiently found for chordal graphs, which makes it the most efficient algorithm for this class of graphs. From all our experiments on different types of general graphs, we conclude that Snowball consistently outperforms Johnson (and Floyd–Warshall), except when the induced width is very high. Our experiments also show that Snowball always outperforms both Chleq–APSP and Snowball–Separators. Although the latter has a better bound on its run time, surprisingly its actual performance is worse than Snowball on all instances of our benchmark sets. This holds even for those instances for which Snowball–Separators performs significantly fewer updates. Thus, we conclude that the additional bookkeeping required by Snowball–Separators does not pay off.

Regarding these experiments, it must be noted that, although we did the utmost to obtain a fair comparison, a constant factor in the measurements depends in a significant way on the exact implementation details (e.g. whether a lookup-table or a heap is used), as is also put forward in earlier work on experimentally comparing shortest path algorithms (Mondou, Crainic, & Nguyen, 1991; Cherkassky, Goldberg, & Radzik, 1996). In our own implementation a higher constant factor for the Snowball algorithms may be caused by adhering to the object-oriented paradigm, i.e. inheriting from the DPC and P³C superclasses, and choosing to reuse code rather than inlining method calls. Nonetheless, we are confident that the general trends we identified hold independently of such details.

Note that strictly speaking, the algorithms introduced in this paper compute all-pairs shortest *distances*. If one wants to actually trace shortest paths, the algorithms can be extended to keep track of the midpoint whenever the distance matrix is updated, just like one does for Floyd–Warshall. Then, for any pair of vertices, the actual shortest path in the graph can be traced in $\mathcal{O}(n)$ time.

In our current implementation of Snowball–Separators, we used a priority queue to decide heuristically which clique tree node to visit next, giving precedence to nodes connected by a large separator to the part of the clique tree already visited. As noted before, we defer answering the question whether an optimal ordering can be found efficiently to future work. We remark that using the minimum-degree heuristic for triangulation provides Snowball with a natural edge, delaying the processing of vertices where the number of iterations of the middle loop is small until $k$ grows large.

Cherkassky and Goldberg (1999) compared several innovative algorithms for *single-source* shortest paths that gave better efficiency than the standard Bellman–Ford algorithm in practice, while having the same worst-case bound of $\mathcal{O}(nm)$ on the run time. In future work, we will investigate if any of these clever improvements can also be exploited in Snowball.

Snowball–Separators can be improved further in a way that does not influence the theoretical complexity but may yield better performance in practice. Iterating over $V_{\text{other}}$ can be seen as a reverse traversal of the part of the clique tree visited before, starting at $c$'s parent. Then, instead of always using the separator between the current clique node (containing $k$) and its parent for all previously visited vertices in $V_{\text{other}}$, we can keep track of the smallest separator encountered during this backwards traversal for no extra asymptotic cost. Since it was shown in Table 1 that the *largest* minimal separator is often hardly smaller than the induced width, it might well pay off to search for smaller separators. We plan to implement this improvement in the near future.

Another possible improvement is suggested by the following observation on DPC. A variant of DPC can be proposed where edge directionality is taken into account: during iteration $k$, only those neighbours $i, j < k$ are considered for which there is a directed path $i \to k \to j$, resulting in the addition of the arc $i \to j$. This set of added arcs would often be much smaller than twice the number of edges added by the standard DPC algorithm, and while the graph produced by the directed variant would not be chordal, the correctness of Snowball would not be impacted.

Furthermore, we would like to also experimentally compare our algorithms to the recent algorithms by Pettie (2004) and the algorithms for graphs of constant treewidth by Chaudhuri and Zaroliagis (2000) in future work. In addition, we are interested in more efficient triangulation heuristics, or triangulation heuristics with a guaranteed quality, to be able to give a guaranteed theoretical bound on general graphs. Another direction, especially interesting in the context of planning and scheduling, is to use the ideas presented here to design a faster algorithm for dynamic all-pairs shortest paths: maintaining shortest paths under edge deletions (or relaxations) and additions (or tightenings).

## Acknowledgments

Roman van der Krogt is supported by Science Foundation Ireland under Grant number 08/RFP/CMS1711.

We offer our sincere gratitude to our reviewers for their comments, which helped us improve the clarity of the article and strengthen our empirical results.

This article is based on a conference paper with the same title, which has received an honourable mention for best student paper at the International Conference on Automated Planning and Scheduling (Planken, de Weerdt, & van der Krogt, 2011).

## Appendix A. Johnson's Heap

In the experiments in this paper, we presented the results for Johnson using a Fibonacci heap, because only then the theoretical bound of $\mathcal{O}\left(nm + n^2 \log n\right)$ time is attained. In practice, using a binary heap for a theoretical bound of $\mathcal{O}\left(nm \log n\right)$ time turns out to be more efficient on some occasions, as we show by the results in this section.

Figure 16 shows the run times of Johnson with a binary heap and with a Fibonacci heap on all of the benchmark sets listed in Table 1. On the diamonds, HTN, and New York benchmarks the binary heap is a few percent faster than the Fibonacci heap, but the slope of the lines in this doubly logarithmic scale is the same, so we can conclude that the average-case run time has similar asymptotic behavior. However, for larger job-shop problems, a binary heap is a factor 2 slower than a Fibonacci heap, and on our chordal graph benchmark problems even a factor 10. Our benchmark problems on scale-free graphs with a fixed number of vertices help explaining this difference.

In Figure 17, the run time of both variants of Johnson can be found for scale-free graphs with 1,000 vertices, with the number of edges varying from about 2,000 to almost 80,000. Here, we see that only for the sparsest scale-free graphs with about 2,000 edges, the binary heap is slightly faster, but when more edges are considered, using the Fibonacci heap significantly outperforms using the binary heap. In particular, the run time of the Fibonacci heap implementation increases only slowly with the number of edges, while the run time of the binary heap increases much more significantly. This can be explained by the fact that when running Dijkstra's algorithm as a subroutine in Johnson, each update of a (candidate) shortest path can be done in amortized constant time with a Fibonacci heap, while in a binary heap this has a worst-case cost of $\mathcal{O}\left(\log n\right)$ time per update. The number of updates is bounded by $m$ for each run of Dijkstra's algorithm, yielding a bound of $\mathcal{O}\left(nm\right)$ updates for Johnson. For the binary heap this $\mathcal{O}\left(nm \log n\right)$ bound accounts for a significant part of the run time, while with a Fibonacci heap other operations (such as extracting the minimum element from the heap) have a bigger relative contribution to the run time.

Based on the results over all benchmark sets, we conclude that although Johnson with a binary heap can help reducing the actual run time in sparse graphs, Johnson with a Fibonacci heap is overall the better choice if $m$ can be large.

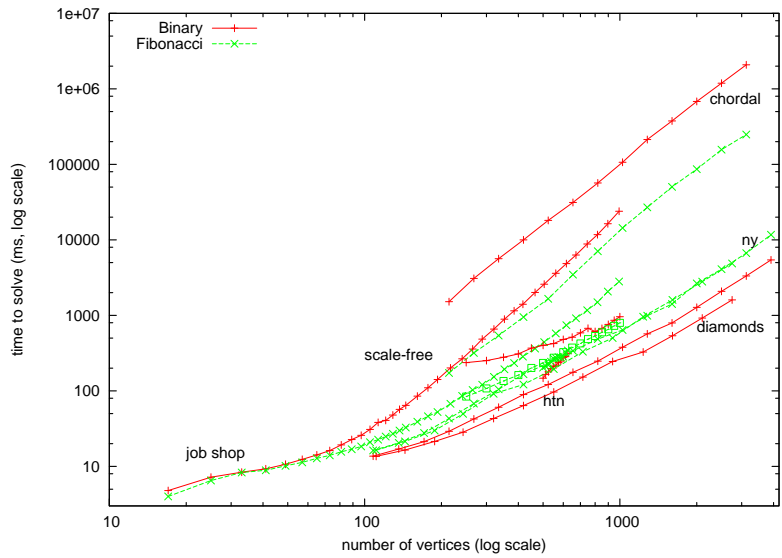

Figure 16: Run times of `Johnson` with a binary heap and with a Fibonacci heap on the benchmark problem sets listed in Table 1.

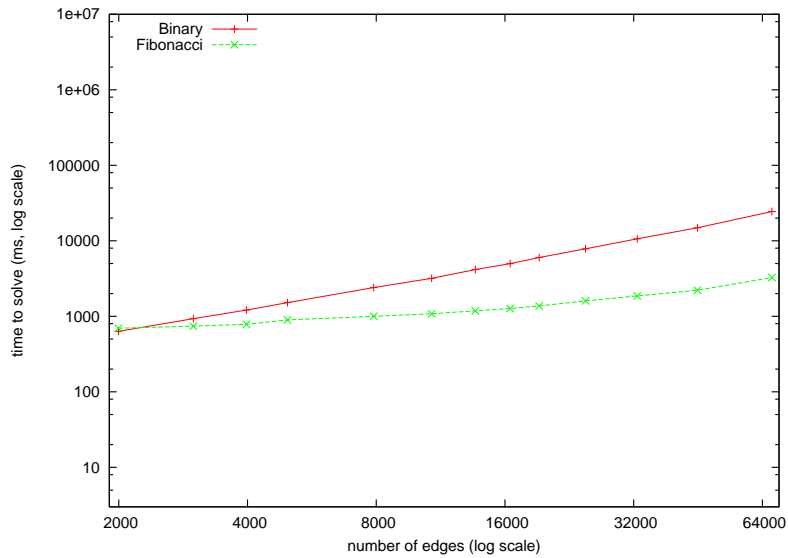

Figure 17: Run times of `Johnson` with a binary heap and with a Fibonacci heap on scale-free graphs with 1,000 vertices and increasing number of edges.